\documentclass[pra,aps,amsmath,amssymb,onecolumn,floatfix,superscriptaddress]{revtex4-2}
\usepackage{amsmath}
\usepackage{graphicx}
\usepackage{color}
\usepackage{amssymb}
\usepackage{soul}
\usepackage[utf8]{inputenc}
\usepackage[T1]{fontenc}
\usepackage{array}
\begin{document}
\title{Effect of beyond mean-field interaction on the structure and dynamics of the one-dimensional quantum droplet}
\author{Sonali Gangwar}
\affiliation{Department of Physics, Indian Institute of Technology, Guwahati 781039, Assam, India} 
\author{Rajamanickam Ravisankar}
\affiliation{Department of Physics, Indian Institute of Technology, Guwahati 781039, Assam, India} 
\affiliation{Institute of Atomic and Molecular Sciences, Academia Sinica, Taipei 10617, Taiwan}
\author{Paulsamy Muruganandam}
\affiliation{Department of Physics, Bharathidasan University, Tiruchirappalli 620024, Tamilnadu, India}
\author{Pankaj Kumar Mishra}
\affiliation{Department of Physics, Indian Institute of Technology, Guwahati 781039, Assam, India}

\date{\today}

\begin{abstract}
We present simulation results of the ground state structure and dynamics of quantum droplets in one-dimensional spin-orbit coupled binary Bose-Einstein condensates. We have considered two cases for this analysis, such as (i) the mean-field term has a vanishingly small contribution utilizing the equal and opposite inter- and intraspecies interaction and (ii) unequal inter- and intraspecies interaction. For both cases, it shows remarkably different natures of the quantum droplet. In the former case, it exhibits bright sech-like droplet nature, while in the latter case, we find the flattened sech-like  shape of the droplet. Further, we analyze the effect of velocity perturbation on the dynamics in both cases. For the first case, we find a systematic change from the solitonic droplet nature to the breathing droplet which finally has a moving droplet feature upon increasing the velocity. However, the second case shows similar dynamics except having more dynamically stable features than the first. Finally, we present various dynamics that ensued in the quantum droplet due to the quenching of the interaction parameters, coupling parameters or allowing the droplet to undergo collisions.
\end{abstract}

\flushbottom

\maketitle

\section{Introduction}
After the realization of Bose-Einstein Condensates (BECs) in the laboratory experiment, the field of ultracold gases has witnessed unprecedented growth in exploring the wealth of novel phenomena in the last few decades, like, supersolid~\cite{Li2017, Baillie2018}, Anderson localization~\cite{Roati2008}, Polarons~\cite{Compagno2017}, self-bound quantum droplet state~\cite{Luo2021}, etc. One of the rapidly developing topics of intense research is the realization of the quantum droplet (QD) state in the condensate. Self-bound QD is a result of the balance between the attractive mean-field interaction and the repulsive beyond mean-field (BMF) interaction owing to the quantum fluctuation, also popularized as Lee-Huang-Yang (LHY) correction~\cite{Lee1957, Skov2021}. Using the detailed behaviour of the collective excitation spectrum, Petrov in 2015 proposed that we could stabilize the attractive collapse of the condensate by the quantum fluctuation, which leads to the formation of a localized state in a binary mixture of two different atoms or different hyperfine states of the same atom~\cite{Petrov2015}. The behaviour of BMF terms depends upon the dimension. It is attractive in one dimension while it becomes repulsive in two and three dimensions~\cite{Petrov2015, Petrov2016, Luo2021}. 

Following the theoretical demonstration of the formation of QD and the crucial role played in this by beyond mean field LHY correction by Petrov, the research on the QD has picked up unprecedented growth in the theoretical, experimental, and numerical directions~\cite{Luo2021}. Post Petrov work, the first QD was observed experimentally with dipolar dysprosium ($^{164}$Dy) condensates with an elongated shape of droplet~\cite{Barbut2016, Schmitt2016, Barbut2016b}. Similar kinds of droplets were also realized for the condensate of erbium atoms ($^{166}$Er)~\cite{Chomaz2016}. Subsequently, many groups attempted to achieve the QD in different condensate systems such as in binary mixtures of isotropic BECs~\cite{Cabrera2018, Cheiney2018, Semeghini2018, Ferioli2019}, heteronuclear bosonic mixtures~\cite{Derrico2019}, Bose-Fermi mixture~\cite{Rakshit:2019} and also in binary magnetic gases~\cite{Smith:2021}. These quantum droplets are commonly observed in three dimensions, while it is possible to increase the lifetime of the droplets in the lower dimensions~\cite{Tylutki2020}. The formation of one-dimensional quantum droplets happens due to the balance between the repulsive mean-field (MF) contribution in energy per particle, which is linear in the density ($n$) of the gas, and the attractive BMF correction, proportional to $-n^{1/2}$~\cite{Petrov2016}. The prediction of the non-trivial attractive nature of the LHY term in quasi-one dimensional quantum droplet~\cite{Petrov2016} has led a significant number of experimental~\cite{Frolian2022}, as well as theoretical and numerical works, in recent years using the effective one component~\cite{Astrakharchik2018}, binary~\cite{Mistakidis2021} and Spin-orbit (SO) and Rabi coupled binary BECs~\cite{Tononi2019, Gangwar2022}. These studies mostly explored the structure and dynamics of the QD.

In general, subject to the mean-field contribution along with BMF, the condensate exhibits two types of self-bound states, such as a bright soliton with zero mean-field~\cite{Jorgensen2018} and a flat-top plateau droplet with a finite mean-field in one dimension. Depending upon the number of particles and interaction strength, Cheiney et al. experimentally observed the presence of these two self-bound states and explored a possible mechanism for this transition~\cite{Cheiney2018}. It exhibits a soliton-like ground state for a small number of atoms $N$, while for large $N$, a flat top plateau droplet was realized. Following this, several numerical simulations were performed using the mean field Gross-Pitaevskii equations (GPEs) with either the effective one-component BEC or binary BECs. Astrakharchik and Malomed considered the effective one-component GPEs with finite mean-field contribution and identified the presence of Gaussian-like droplet for small $N$ and flat top droplet for large $N$~\cite{Astrakharchik2018}. However, Mistakidis et al. reported a similar type of QD in binary BECs in one dimension~\cite{Mistakidis2021}. In a similar line, Li et al. reported the presence of stable QD with attractive intercomponent and repulsive intracomponent interaction for 2D SO coupled binary BECs with LHY and analyzed the effect of SO coupling terms on the stability of the droplet~\cite{Li2017A}. Some works highlight the effect of BMF on the structure and dynamics of the QD for vanishingly small mean-field contribution~\cite{Tononi2019, Gangwar2022}. Tononi et al., in SO coupled BECs, demonstrated that the self-bound states are solitonic for vanishingly small mean-field contribution~\cite{Tononi2019}. Depending upon the Rabi and SO coupling parameters range, these soliton-like states are either of single peak (bright soliton) or multiple peaks (stripe soliton) nature~\cite{Chiquillo2018, Tononi2019, Gangwar2022, Vinayagam2017}. 

Apart from the structural analysis of the QD, the main thrust has been to analyze the dynamical robustness of these self-bound droplet states. The dynamics can be attained through several means such as, by giving initial velocity to the condensate~\cite{Tononi2019, Gangwar2022}, quenching the coupling parameters~\cite{Ravisankar2020}, and quenching the interaction strength~\cite{Mistakidis2021}. Mistakidis et al. demonstrated the appearance of the expanding and breather-like droplets upon quenching the interspecies interaction~\cite{Mistakidis2021}. Edmonds et al. showed the transition of the single droplet state into multiple one along with atomic radiation and bright soliton-like excitations upon quenching the interaction strength for the dipolar droplet~\cite{Edmonds2020}. Apart from investigating the dynamics of droplets, researchers have also tried to draw a one-to-one parallel comparison of the collisional dynamics of the QD in a similar line of classical droplets. In general, the collision between droplets yields two peculiar behaviour, namely the droplets get merged into a single one and thus undergo inelastic collision in the first case, and they undergo elastic collisions in the second case~\cite{Pan2008}. Depending upon the velocity of the droplets, Ferioli et al. experimentally demonstrated similar nature of the collision between the QD in the binary mixture of $^{39}$K ~\cite{Ferioli2019}. At low velocities, droplets undergo inelastic collision, while they have an elastic collision at higher velocities. Similar collisional dynamics for the QD have also been reported numerically for binary BECs in 1D, with~\cite{Pathak2022} or without trap~\cite{Astrakharchik2018}. 

From the theoretical and numerical front, the researchers have resorted to the effective one-component BECs model to analyze the structure and dynamics of the QD. For instance, Astrakharchik and Malomed considered the effective one-component dynamics by taking the symmetric nature of the binary BECs~\cite{Astrakharchik2018}. This assumption imposes a limitation while exploring the dynamics of individual components of the QD~\cite{Tononi2019, Gangwar2022}. Recently we have shown that in the presence of the vanishingly mean-field contribution, the ground state of SO coupled BECs with the BMF term exhibits either quantum soliton or quantum stripe soliton. We demonstrated a systematic transition from the soliton into the breathing soliton, which further makes a transition into the moving soliton upon increasing the velocity perturbation in the ground state of the self-bound state. We also reported a large variety of novel features like the dynamical phase transition from plane to the stripe soliton, dynamical flipping of the spin state, generation of the secondary solitons, etc., when the dynamics are generated through the quenching of the coupling parameters. However, in the presence of finite mean-field terms, the dynamical robustness of the quantum soliton has not been investigated in great detail for the SO coupled BECs~\cite{Gangwar2022}. In this paper, we consider the effect of finite mean-field interaction on the overall structure and dynamics of the QD and show the transition from the quantum bright soliton to the quantum droplet (with flat top) state upon the increase in the mean-field contribution by keeping the LHY contribution fixed. We demonstrate a significant role played by the mean-field contribution to the nature of the collision between the droplet. The collision is inelastic for vanishingly small mean-field contributions, while for a higher value of the mean-field interaction, it is elastic in nature. 

The structure of our paper is as follows. In Sec.~\ref{sec:2}, we present governing equations and numerical simulation details. Following this in Sec.~\ref{sec:3}, we provide a detailed analysis of the ground state structure for a vanishingly small mean-field term and finite mean-field contribution followed by their dynamics, which have been set up in the system through different means. First, we present the different sorts of dynamics that appear in the system due to different velocities, followed by the dynamics due to the quenching of the nonlinear interactions. Further, we present some of the interesting dynamical behaviours that appear due to collision for both finite and zero mean field terms. Finally, we conclude our work in Sec.~\ref{sec:4}.

\section{Beyond Mean-field model for SO coupled BEC\lowercase{s}}
\label{sec:2}
We consider the one-dimensional pseudo-spin-$1/2$ spinor condensate in strongly transverse confined traps. The corresponding coupled GPEs in dimensionless form can be written as~\cite{Tononi2019,Gangwar2022}:
\begin{subequations}
\label{eq:gpsoc:1}
\begin{align}
\mathrm{i} \partial_t \psi_{\uparrow}=\bigg[ &-\frac{1}{2}\partial_x^2-\mathrm{i} k_{L} \partial_x + \frac{\delta g}{2}\left(\lvert \psi_{\uparrow} \rvert^2 + \lvert \psi_{\downarrow} \rvert^2  \right) 
+ g \lvert \psi_{\uparrow} \rvert^2 + g_{\uparrow\downarrow} \lvert \psi_{\downarrow} \rvert^2 
-\dfrac{g_{\text{LHY}}^{3/2}}{\pi}\sqrt{\lvert \psi_{\uparrow}\rvert^2 +\lvert\psi_{\downarrow}\rvert^2}\bigg] \psi_{\uparrow}+ \Omega \psi_{\downarrow}, \label{eq:gpsoc2-a} \\
\mathrm{i} \partial_t \psi_{\downarrow}=\bigg[ &-\frac{1}{2}\partial_x^2 +\mathrm{i} k_{L} \partial_x  
+ \frac{\delta g}{2}\left(\lvert \psi_{\uparrow} \rvert^2 + \lvert \psi_{\downarrow} \rvert^2  \right) 
+ g_{\downarrow\uparrow} \lvert \psi_{\uparrow} \rvert^2 + g \lvert \psi_{\downarrow} \rvert^2 -\dfrac{g_{\text{LHY}}^{3/2}}{\pi}\sqrt{\lvert \psi_{\uparrow}\rvert^2 +\lvert\psi_{\downarrow}\rvert^2}\bigg] \psi_{\downarrow}+ \Omega \psi_{\uparrow}, \label{eq:gpsoc2-b}
\end{align}
\end{subequations}
where $\psi_\uparrow$ and $ \psi_\downarrow$ correspond to the wavefunctions of the spin-up and spin-down components, respectively, $k_L$ is the spin-orbit coupling strength, $\Omega$ is the Rabi coupling frequency, $g$ accounts the intraspecies interaction, and $g_{\uparrow\downarrow}$ represents interspecies interaction strength. $g_{\text{LHY}}=g$ is the interaction due to LHY correction, and $\delta g = g_{\uparrow \downarrow} + g$ represents the imbalance between the mean-field contributions from the inter and intra-species interactions. Note that for $g=-g_{\uparrow \downarrow}$ implies $\delta g=0$, which has been used to analyze the self-bound state (quantum soliton) for the spin-orbit coupled BECs~\cite{Gangwar2022}. In this paper, we have considered the different ranges of $\delta g/g$, which yields the finite contribution from the overall mean-field term in the self-bound state. The normalization of the condensate wave function follows~\cite{Sakaguchi2014,Cheng2014}:

\begin{align}
 N_{total} = \int\limits_{-\infty}^{\infty} 
 ( \lvert \psi_\uparrow \rvert^2 + \lvert \psi_\downarrow \rvert^2 ) \, dx \equiv N_{\uparrow} + N_{\downarrow},
\end{align}
\color{black}
which remains invariant with time, here $N_{\uparrow} = N_{\downarrow} = N$. 

We have chosen transverse harmonic oscillator length $a_{0}=\sqrt{\hbar/(m\omega_{\perp})}$ as a characteristic length scale (where, $\omega_\perp$ is the trap frequency in the transverse direction), $\omega_{\perp}^{-1}$ as a time scale, and $\hbar\omega_\perp$ as an energy scale to obtain the non-dimensionalize Eqs. (\ref{eq:gpsoc:1}). Here $g$ = $2 N a_{\uparrow \uparrow}/ a_{0}$ and $g_{\uparrow\downarrow} = 2 N a_{\uparrow \downarrow} / a_{0}$ with $N$ as number of particles, $a_{\uparrow \uparrow}$ and $a_{\uparrow \downarrow}$ being the scattering length corresponding to intra- and inter-species components, respectively. The corresponding adimensional scheme for the SO and Rabi coupling is $k_L \to k_L a_{0}$ and $\Omega \to {\Omega}/\omega_{\perp}$. The wave function is rescaled as $ \psi_{\uparrow, \downarrow} = \psi_{\uparrow, \downarrow} \sqrt{a_{0}}$.

The ground state is obtained by numerically solving the coupled GP equations~(\ref{eq:gpsoc2-a}) and (\ref{eq:gpsoc2-b}) using the imaginary time method with the aid of split-step Crank-Nicolson scheme~\cite{Gangwar2022, Ravisankar2021}. We consider the box size for all the simulations as $[-153.6: 153.6]$ with a spatial resolution of $dx = 0.025$. We assume the Gaussian initial condition with anti-symmetric profiles on the components, i.e., ${\psi}_{\uparrow}(x)=-{\psi}_{\downarrow}(-x)$ to obtain all the ground states. For all the simulations, the time step is fixed at $dt = 10^{-5}$. We consider the interaction strength as $g=0.5$ and take $\delta g/g=0,0.1,0.5$.

\section{Numerical results}
\label{sec:3}
In the recent numerical studies, we showed that for the vanishingly small mean-field contribution ($\delta g=0$), depending upon the coupling parameters range, we may observe either a quantum-bright droplet (QBD) or quantum stripe droplet (QSD) for SO coupled BECs with LHY correction~\cite{Gangwar2022}. However, in the laboratory experiment, $\delta g$ happens to be finite~\cite{Frolian2022}. Also, some recent theoretical and experimental observations report the transition from the Gaussian-like droplet to the flat top-like droplets either upon increasing the $\delta g/g$ or increasing the number of particles~\cite{Mithun2020, Mistakidis2021}. In continuation of our previous work on the structure of the droplets for $\delta g=0$~\cite{Gangwar2022}, in this section, we present the effect of finite but small $\delta g/g$ on the ground state phases of the SO coupled droplet. Further, we provide the dynamics of the droplet for different sorts of perturbations, namely by giving initial equal and opposite velocities to the components, quenching the interaction parameters and allowing the condensate component to collide in the presence of $\delta g/g$ and analyzing the dynamics of the droplet for $\delta g/g \ne 0$.
\subsection{Structure and dynamics of the droplet}
\label{sec:3a}
\begin{figure}[!htp]
\centering\includegraphics[width=0.99\linewidth]{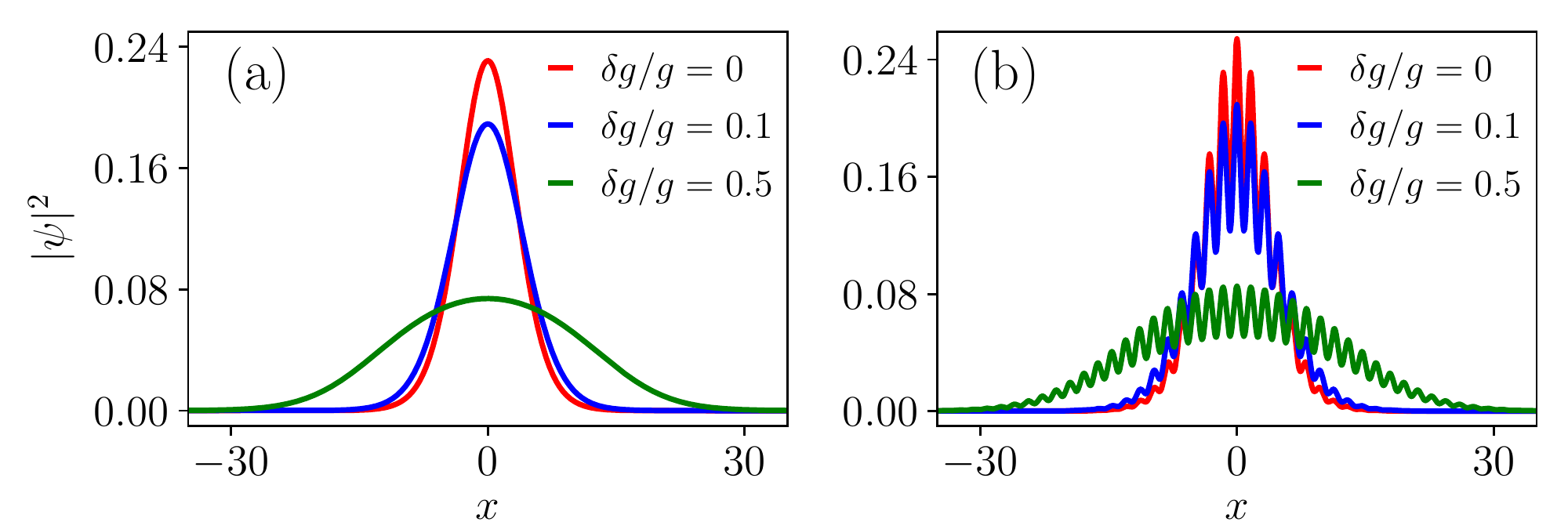} 
\caption{Ground state density profiles for (a) quantum bright droplet phase ($\Omega = 1$ and $k_L=0.5$) and (b) quantum stripe droplet phase ($\Omega = 1$ and $k_L=2$) for different $\delta g/g$. Upon increasing the $\delta g/g$, the peak density reduces, and the profile gets flattened in both phases.}
\label{fig:denqs}
\end{figure}


We now demonstrate the structure of the droplet by considering the deviation from vanishingly small mean-field term ($\delta g =0$). To make the $\delta g/g$ finite, however, $\delta g/g <1$, one of necessary arrangment used for the droplet formation~\cite{Petrov2016}, we have varied the interspecies interaction keeping intraspecies interaction fixed. We first consider the quantum-bright droplets, which we prepared by assuming $g=0.5$, $\Omega=1$, $k_L=0.5$, and $g_{\uparrow\downarrow}=-0.5, -0.45, -0.25$ and employing $\delta g/g =0, 0.1, 0.5$, respectively. In Fig.~\ref{fig:denqs}(a), we show the ground states for different $\delta g/g$ ($\delta g/g=0,0.1,0.5$). We find that the ground state for the finite $\delta g/g$ shows QBD the same as those obtained for $\delta g/g=0$. Upon increasing the $\delta g/g$ we notice a decrease in the density peak along with an increment in the shape of the droplets that leads a transformation of the droplet from the sech-like shape to the broader sech-like~\cite{Petrov2016, Mistakidis2021}. Fig.~\ref{fig:denqs}(b) depicts the size of the quantum stripe droplets for different $\delta g/g$ with $\Omega =1$ and $k_L =2$. Similar to the change in shape and size of the QBD phase droplets upon the increase of $\delta g/g$, here we also find that the peak density of QD decreases upon increasing $\delta g/g$, while the size of the droplet gets bigger. The overall features of the droplets upon increasing $\delta g/g$ become consistent with those reported for the binary BECs~\cite{Mistakidis2021}.

Next we focus on analyzing different sorts of dynamics of the droplet. We have attained the dynamics using the three protocols, namely, by giving a small initial velocity to each component, making a sudden quench in the nonlinear interaction parameters, and allowing the collision between the droplets. In what follows, we analyze these dynamics for $\delta g/g=0$ and $\delta g/g \ne 0$. The dynamics of the ground state of quantum droplets have been explored by numerically solving the GPEs [cf. Eqs.~(\ref{eq:gpsoc:1})] employing the real-time propagation scheme~\cite{Gangwar2022}. First, we explore the dynamics by providing initial velocity to the condensate. Depending on the magnitude of the velocity, we obtain the appearance of different sorts of dynamics of the droplet, like breathing droplets, fragmented droplets, etc.~\cite{Gangwar2022}.



\begin{figure*}[!ht]
\centering\includegraphics[width=0.99\linewidth]{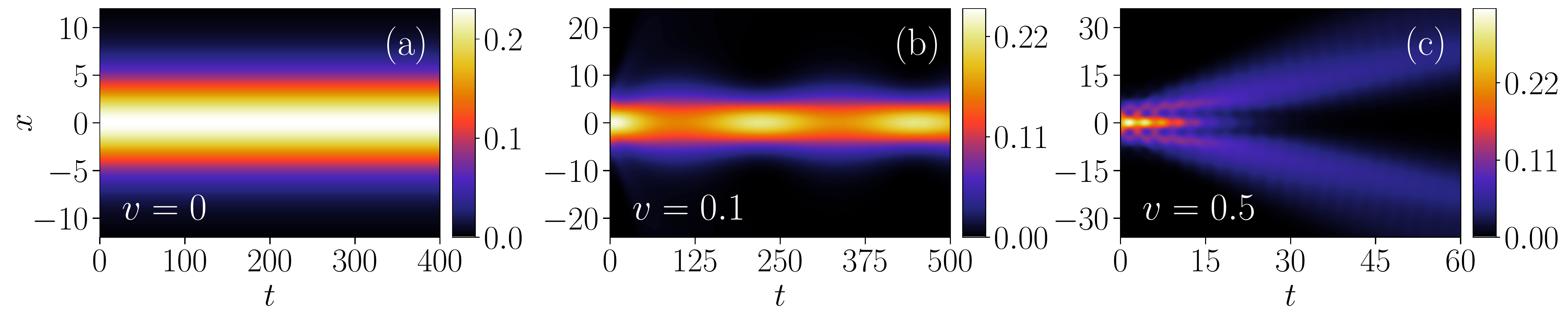} 
\caption{Dynamics of the quantum-bright droplet for different initial velocities for $\delta g/g =0$: (a) $v=0$ (b) $v=0.1$ and (c) $v=0.5$. The ground state was prepared with the interaction strength $g= -g_{\uparrow\downarrow}=0.5$, $\Omega = 1$, and $k_L = 0.5$. For $v=0$, no change in the shape and size of the density profile happens, while at $v=0.1$, we notice a breathing droplet, which transforms into a moving droplet at $v=0.5$.}
\label{fig:dynapw}
\end{figure*}
\begin{figure}[!ht]
\centering\includegraphics[width=0.99\linewidth]{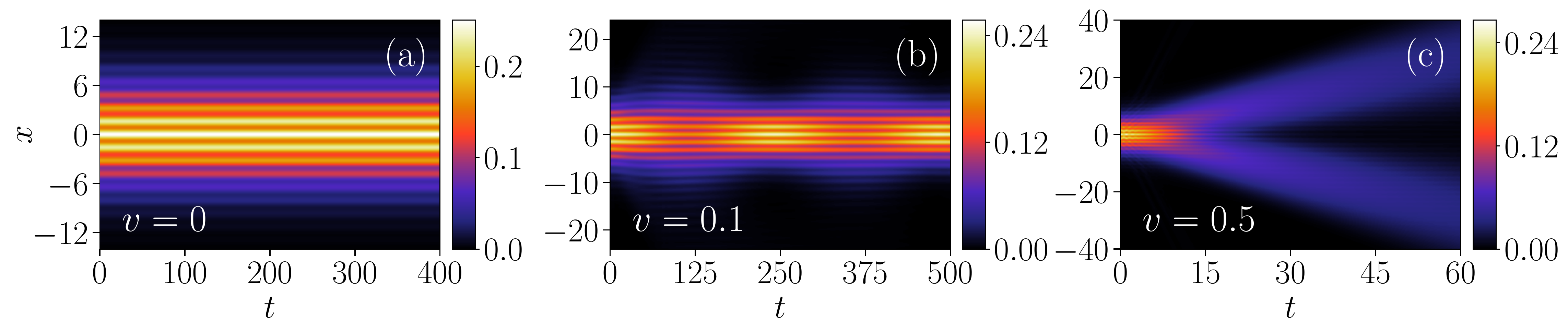} 
\caption{Dynamics of quantum stripe droplet at different initial velocities for $\delta g/g=0$: (a) $v=0$, (b) $v=0.1$, and (c) $v=0.5$. The other parameters are $\Omega =1$ and $k_L = 2$. For $v=0$, the soliton moves without any distortion in shape and size while we observe breathers for $v=0.1$, which further transforms into a moving droplet for $v=0.5$.}
\label{fig:dynasw}
\end{figure}

\begin{figure}[!ht]
\centering\includegraphics[width=0.99\linewidth]{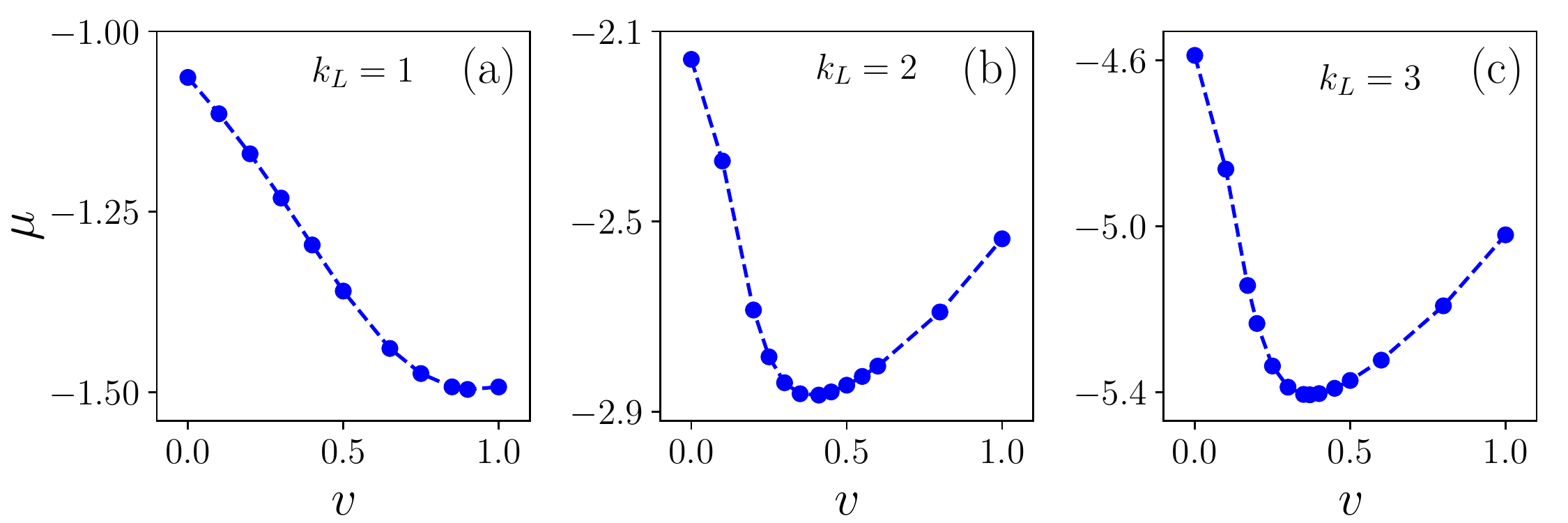} 
\caption{Variation of chemical potential $\mu$ with velocity for different $k_L$: (a)for $k_L=1$, (b) for $k_L=2$, and (c) for $k_L=3$. Other parameters are $\delta g/g=0$ and $\Omega=1$. The velocity at which $\mu$ attains minima provides the critical velocity ($v_c$) at which the breather droplet transforms into a moving droplet. $v_c$ decreases upon increasing the $k_L$.}
\label{fig:chemPot}
\end{figure}

In Fig.~\ref{fig:dynapw}(a-c), we illustrate the temporal evolution of quantum bright droplets at different initial velocities $v=0,0.1$ and $0.5$ as the ground state was prepared for $g = -g_{\uparrow \downarrow}=0.5 (\delta g/g=0)$, $\Omega =1$, and $k_L = 0.5$. Note that the up and down component has been given the velocity in the negative and positive direction respectively~\cite{Gangwar2022}. For $v=0$, we do not find any change in the shape and size of the condensate as the system evolved with time [see Fig.~\ref{fig:dynapw}(a)], indicating the solitonic nature of the droplet. However, for the finite but small velocity ($v=0.1$), the droplet experiences a breathers-like excitation as shown in Fig.~\ref{fig:dynapw}(b). Upon further increasing the $v$ to $0.5$, the droplet undergoes multiple fragmentations and gets split into three droplets as depicted in Fig.~\ref{fig:dynapw}(c). Here the middle droplet remains unmoved, which disappears at later times ($t \sim 20$), and the other two lobes of the fragmented droplet start moving in the positive and negative direction leading to the appearance of moving droplets. Note that the presence of moving droplets is not trivial for the SO coupled BECs, as the SO coupling breaks the Galilean invariance of the system. However, here we find that the moving droplet starts existing beyond a critical velocity. Similar behaviour of the existence of moving solitonic states has been reported for the solitons moving below a critical velocity for two-dimensional SO coupled BECs.~\cite{Sakaguchi2014}. 

In Fig.~\ref{fig:dynasw}, we show the dynamics of quantum stripe droplets for different velocities $v =0, 0.1$ and $0.5$ in which we prepare the ground state for $g=-g_{\uparrow \downarrow} =0.5 (\delta g/g=0)$, $\Omega =1$, and $k_L=2$. For $v=0$, the quantum stripe droplet propagates with no change in the shape and size as shown in Fig.~\ref{fig:dynasw}(a). With a finite velocity $v=0.1$, the stripe phase of the droplet exhibits breather-like expansion and compression [see Fig.~\ref{fig:dynasw}(b)]. At $v=0.5$, similar to the quantum bright droplet stripe phase also has the presence of moving droplets [see Fig.~\ref{fig:dynasw}(c)].

\begin{figure*}[!htp]
\centering\includegraphics[width=0.99\linewidth]{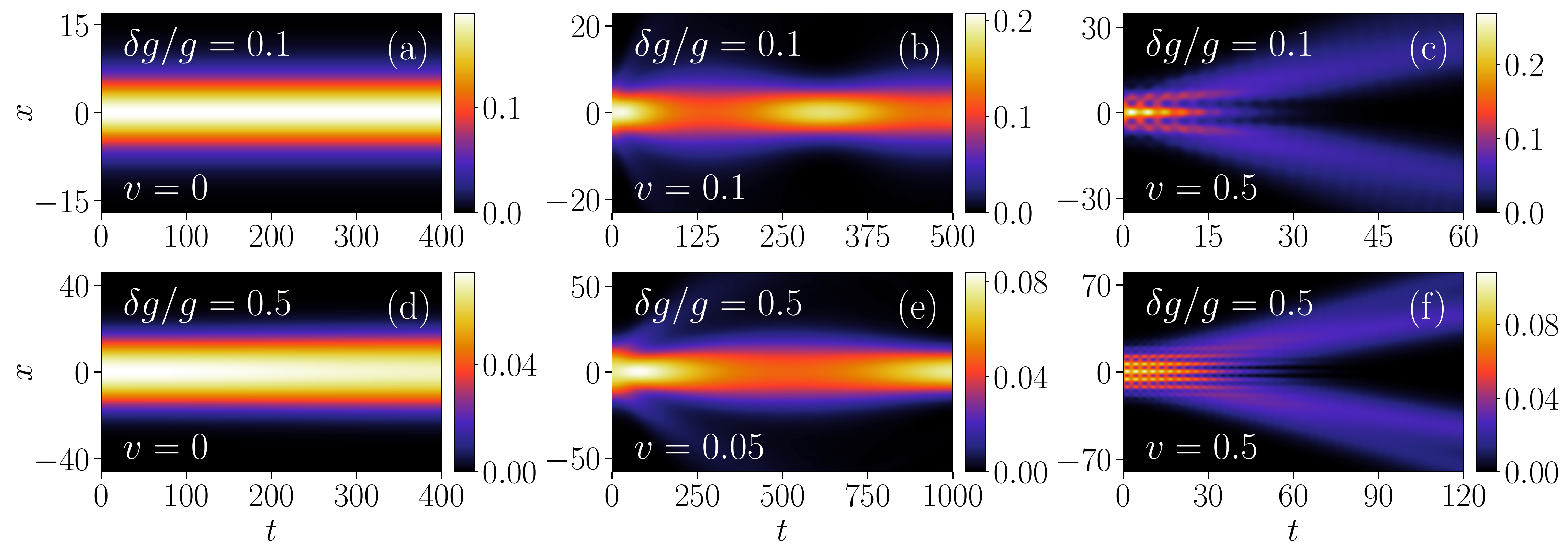} 
\caption{Dynamics of quantum droplet for finite $\delta g/g$ for $\Omega =1$ and $k_L = 0.5$. (a)-(c) $\delta g/g=0.1$ and velocity $v=0, 0.1, 0.5$ respectively. (d)-(f) $\delta g/g=0.5$ for different initial velocities, $v=0, 0.05$ and $0.5$ respectively. For both $\delta g/g$ at $v=0$, the droplet remains stable, while for finite velocity ($v=0.1, 0.05$), we observe a breather droplet, which converts into moving droplets upon increasing the velocity to $v=0.5$. The breathing frequency appears to decrease compared to those for $\delta g/g=0$ for the quantum bright droplet.}
\label{fig:finiteg1}
\end{figure*}

\begin{figure*}[!htp]
\centering\includegraphics[width=1\linewidth]{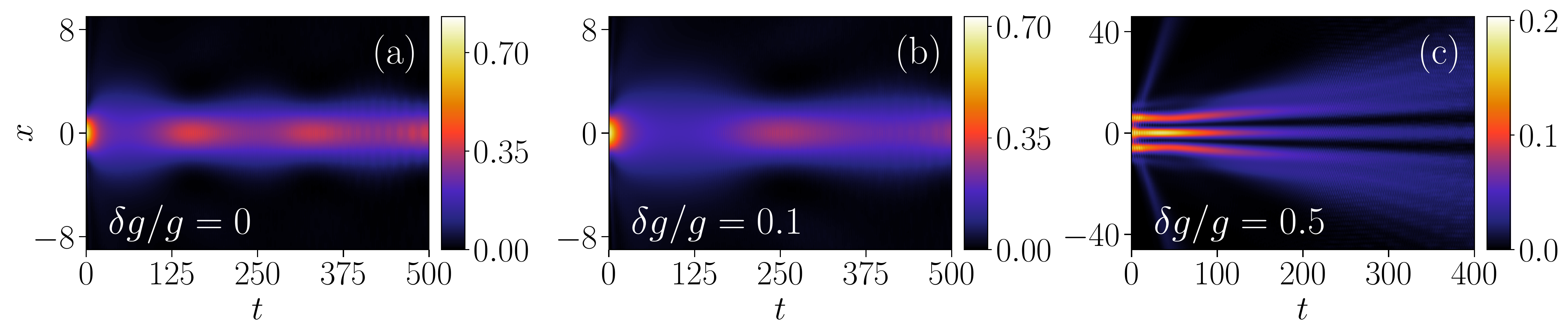} 
\caption{Dynamical evolution of total density of the quantum bright droplet for different $\delta g/g$: (a)$\delta g/g=0$ (b) $\delta g/g= 0.1$ and (c) $\delta g/g= 0.5$ as the initial velocity given to the condensate is $v=0.5$. The other parameters are $\Omega=1, k_L=1$ and $g=0.5$. Upon increase in $\delta g/g$ leads to decrease in the breathing frequency and at high $\delta g/g(\sim 0.3)$ the droplet transforms into the moving droplet.}
\label{fig:dendelgpw}
\end{figure*}

In both the quantum bright and stripe phases of the droplet, we find that the increase in the velocity results in the transition from the breathing droplet to the moving droplet. Astrakharchik and Malomed report a similar feature of the droplet with an increase in the velocity for the binary mixture with an assumption of symmetric behaviour of the components~\cite{Astrakharchik2018}. However, in the present study, we find the presence of Rabi and SO coupling between the components makes these breathing and moving nature of droplet more intricate. To probe the genesis of this peculiar feature, we compute the chemical potentials ($\mu$) corresponding to the bright as well as stripe phase of the droplet and analyze the detailed behaviour. In Fig.~\ref{fig:chemPot}, we show the variation of chemical potential ($\mu$) with the velocity perturbation for different $k_L$ keeping $\Omega=1$ and $\delta g/g=0$ parameters fixed. We find that the chemical potential attains its minimum at the critical velocity ($v_c$) beyond which the droplet exhibits the bifurcation into the moving droplets. For critical velocity of the transition appears to be at $v_c\sim 0.9, 0.41$ and $0.37$ respectively for $k_L=1, 2$ and $3$. The presence of two contrasting different dynamical region of the droplets below and above the critical velocity can be understood as the domination of the breathing droplet region by the SO and Rabi coupling contribution, while the moving droplet region by the domination of the kinetic energy~\cite{Gangwar2022}.

\begin{figure*}[!htp]
 \centering\includegraphics[width=0.5\linewidth]{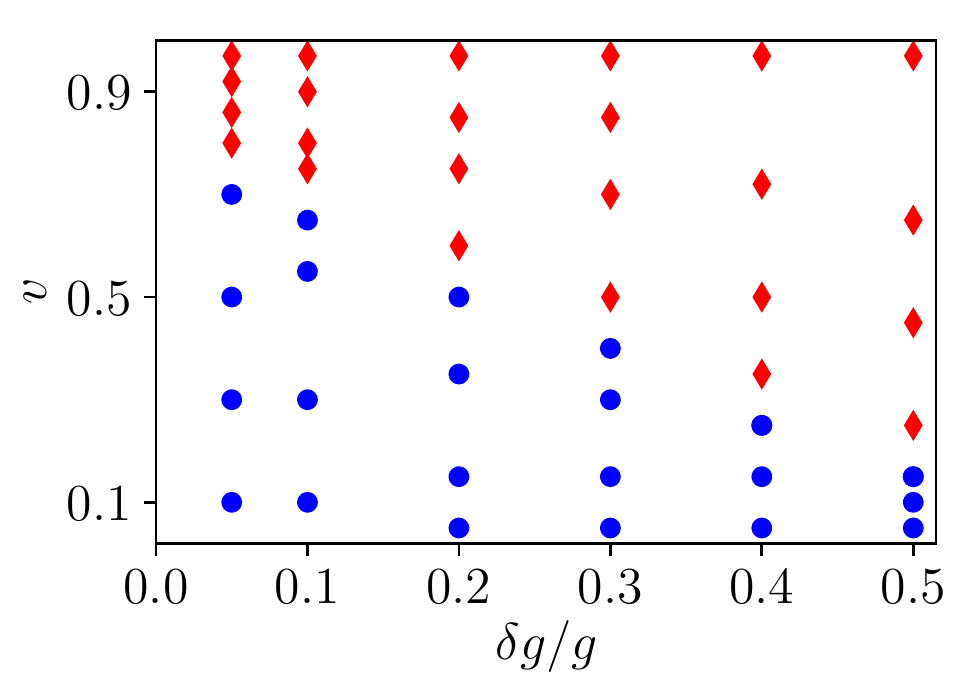} 
\caption{Plot showing the region of the breathing (blue dots) and moving (red diamond) droplets in the parameters space of $v$ and $\delta g/g$ for the bright droplet region. The velocity at which transition from breathing droplet to the moving droplet takes place for a given $\delta g/g$ decreases upon increase in $\delta g /g$.}
\label{fig:phaseplotpw}
\end{figure*}

\begin{figure*}[!htp]
\centering\includegraphics[width=0.99\linewidth]{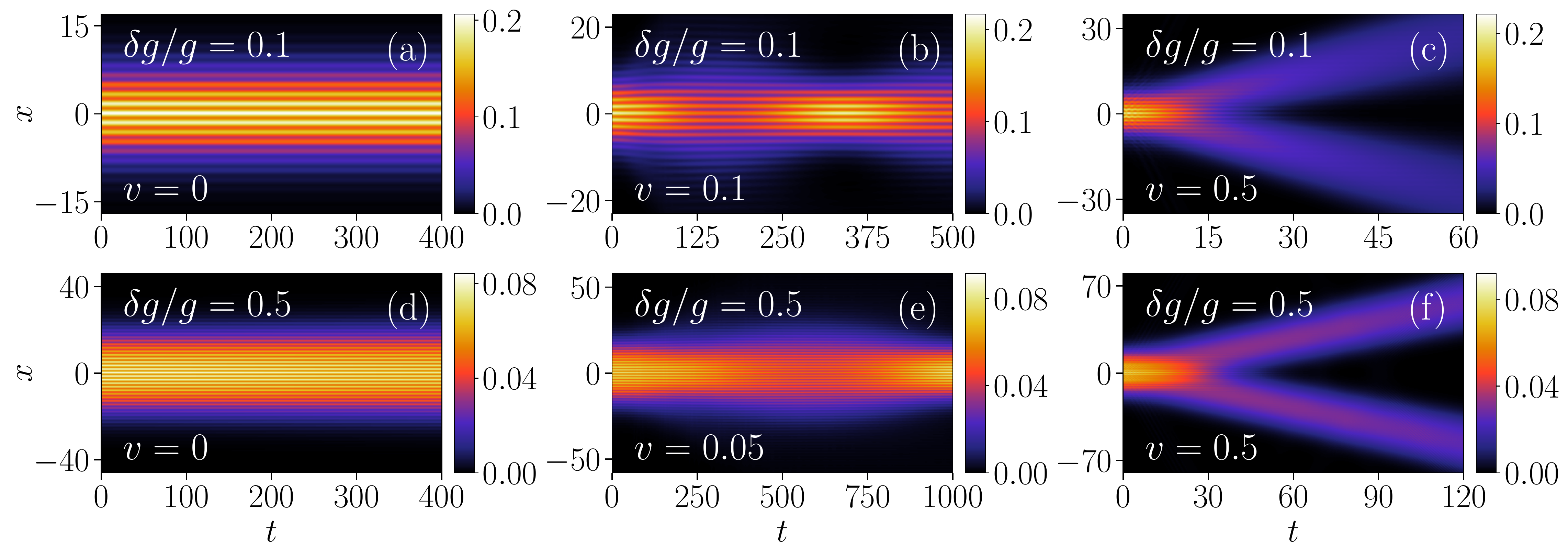} 
\caption{Dynamics of quantum stripe droplet for finite $\delta g/g$ for $\Omega =1$ and $k_L = 2$. (a)-(c): $\delta g/g=0.1$ and velocity $v=0, 0.1, 0.5$ respectively. (d)-(f): $\delta g/g=0.5$ for different initial velocities, $v=0, 0.05$ and $0.5$ respectively. Upon increasing the velocity from $v=0$ to $v=0.1$ for $\delta g/g=0.1$ and from $v=0$ to $v=0.05$ for $\delta g/g=0.5$, we observe a transition from solitonic droplet to breather droplet. For large velocity ($v=0.5$), moving droplets are observed for both $\delta g/g$.}
\label{fig:finiteg2}
\end{figure*}

In Fig.~\ref{fig:finiteg1}, we illustrate the dynamics of quantum bright droplet for finite $\delta g/g$: $\delta g/g =0.1$ (in top panel) and $\delta g/g=0.5$ (bottom panel) for different velocities. In Fig.~\ref{fig:finiteg1}(a)-(c), we fixed the interaction strength $g=0.5$, $g_{\uparrow \downarrow} = -0.45$, while, in Fig.~\ref{fig:finiteg1}(d)-(f) $g=0.5$, $g_{\uparrow \downarrow} =-0.25$, $\Omega=1$, and $k_L=0.5$. For $\delta g/g=0.1$, the droplet exhibits solitonic nature at $v=0$ same as those observed for $\delta g/g =0$. Upon increasing the velocity to $v=0.1$, the droplet exhibits breathing nature as shown in Fig.~\ref{fig:finiteg1}(b). However, the breathing frequency appears to be smaller for finite $\delta g/g$ than those for $\delta g/g=0$ for the same velocity. At large velocity $v=0.5$ droplet transforms into a moving droplet as depicted in Fig.~\ref{fig:finiteg1}(c). Interestingly with finite $\delta g/g$, the bifurcation of moving droplets takes place after a long time compared to those for $\delta g=0$. Further increase in $\delta g/g=0.5$, we find a similar behaviour of the droplet with change in the velocity as those observed for $\delta g/g=0.1$. However, we observe two remarkable differences. First, the frequency of the breathing phase appears to be smaller than those for $\delta g/g=0.1$, and the bifurcation of the droplet happens for an even larger time ($t \sim 25$) than those for $\delta g/g=0.1$ which is $t\sim 10$. Overall, we find that the increase in $\delta g/g$ leads the enhancement in the dynamical stability of the droplet. 

\begin{figure*}[!htp]
\centering\includegraphics[width=1\linewidth]{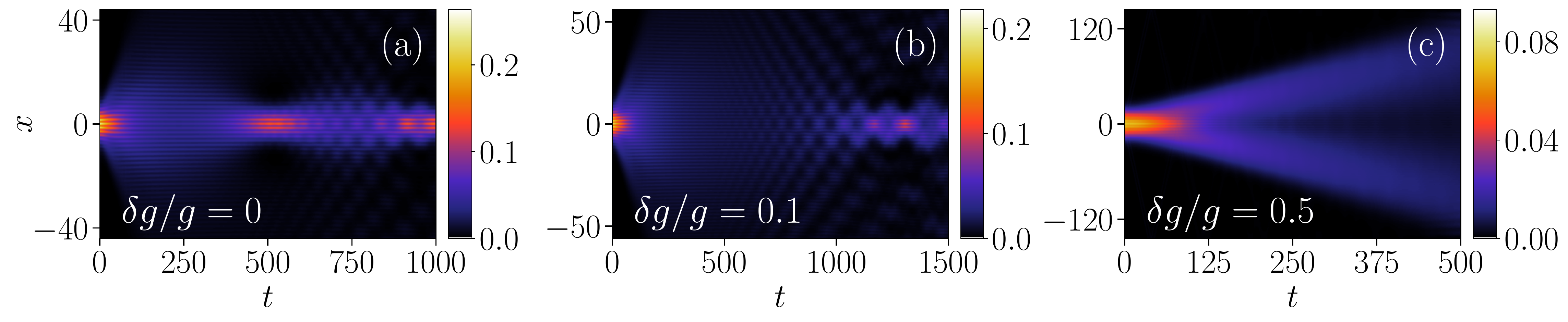} 
\caption{Dynamical evolution of total density of the quantum stripe droplet for different $\delta g/g$: (a)$\delta g/g=0$ (b) $\delta g/g = 0.1$ and (c) $\delta g/g= 0.5$ as the initial velocity given to the condensate is $v=0.2$. The other parameters are same as in Fig.~\ref{fig:finiteg2}. Upon increase in $\delta g/g$ leads to decrease in the breathing frequency and at high $\delta g/g(\sim 0.2)$ the droplet transforms into the moving droplet.}
\label{fig:dendelgsw}
\end{figure*}

In Fig.~\ref{fig:dendelgpw}, we show the dynamical evolution of the total density of the quantum bright droplet by keeping the velocity fixed to $v=0.5$ and varying the $\delta g/g$ keeping $g=0.5$, $\Omega=1$ and $k_L=1$. At $\delta g/g=0$, the droplet exhibits the breathing-like oscillation with time period $T\sim 154$, which decreases to $T\sim 220$ upon the increase of $\delta g/g=0.1$. Upon further increase in $\delta g/g (=0.5)$ results in the appearance of the moving droplets. Overall we find that the transition from the breathing to the moving droplet can be controlled by either tuning the velocity for a fixed mean-field interaction parameter ($\delta g/g$) or by varying the $\delta g/g$ for fixed velocity. In order to get a comprehensive picture of the regime for the breathing and moving droplet we show them in the $\delta g/g$ and $v$ parameter space in Fig.~\ref{fig:phaseplotpw}. The blue dots represent the breathing droplet region while the red diamonds depict the moving droplet region which has been obtained by analyzing the nature of the total density. We find that the increase in $\delta g/g$ makes the transition from breathing to the moving droplet at a relatively lower velocity compared to those for lower $\delta g/g$. This particular feature can be attributed to the increase in the stiffness of the droplet upon the increase in $\delta g/g$~\cite{Astrakharchik2018}. 

\begin{figure}[!ht]
\centering\includegraphics[width=0.99\linewidth]{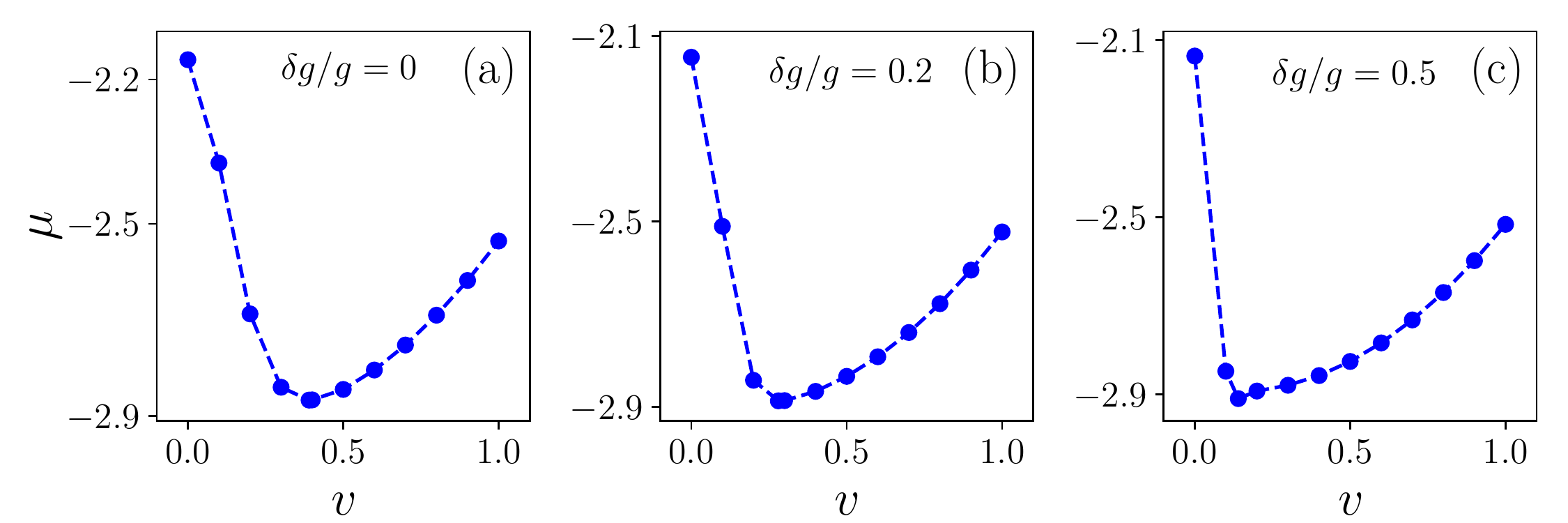} 
\caption{Variation of chemical potential ($\mu$) with velocity for different $\delta g/g$: (a) for $\delta g/g=0$, (b) for $\delta g/g=0.2$, and (c) for $\delta g/g=0.5$ in the stripe droplet region ($\Omega=1$ and $k_L=2$). Other parameters are same as in Fig.~\ref{fig:finiteg2}. The critical velocity at which $\mu$ attains separates the breathing droplet and moving droplet decreases upon increasing $\delta g/g$.}
\label{fig:chemPot1}
\end{figure}

\begin{figure*}[!htp]
\centering\includegraphics[width=0.5\linewidth]{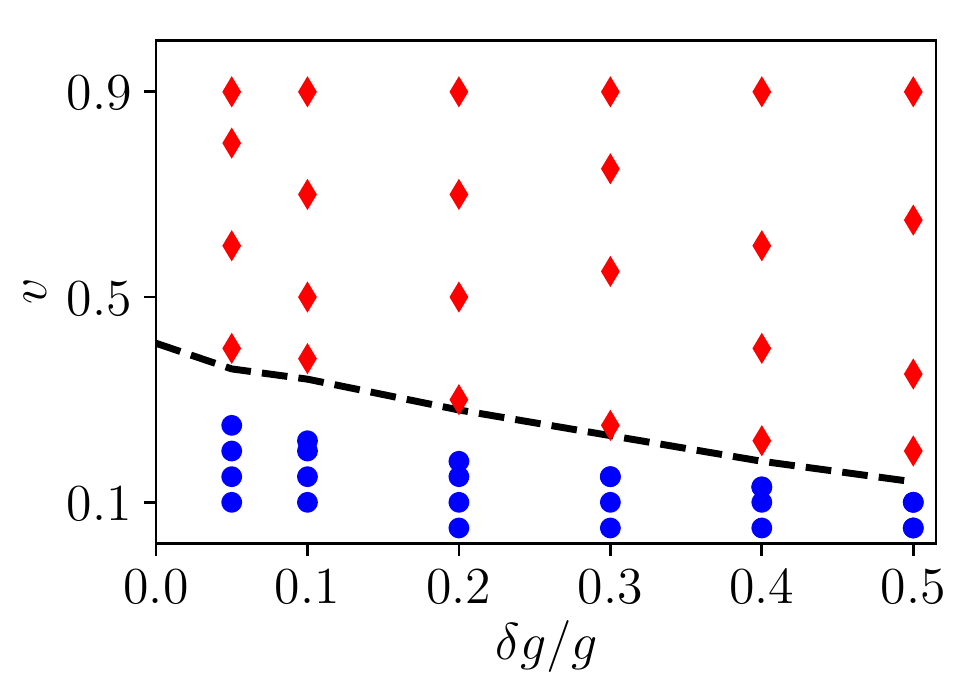} 
\caption{Plot showing the region of the breathing (blue dots) and moving (red diamonds) droplets in the parameters space of $v$ and $\delta g/g$ for the stripe region. The dotted line represents the boundary between the two regimes decided analyzing the points at which the chemical potential attains the minimum. The other parameters are same as in Fig.~\ref{fig:finiteg2}}.
\label{fig:phaseplotsw}
\end{figure*}

The dynamics of the stripe phase have similar nature for different $\delta g/g$ keeping $g=0.5$, $\Omega=1$ and $k_L=2$ as those for the quantum bright phase, as discussed above. In Fig.~\ref{fig:finiteg2}(a)-(c), we show the dynamical evolution of the stripe phase of the droplet for $\delta g/g=0.1$ for the different velocity $v=0, 0.1$, and $0.5$ respectively. We observe a solitonic nature of the droplet for $v=0$, a breather for $v=0.1$ and a moving droplet phase for $v=0.5$. Similarly, with $\delta g/g=0.5$, we also observe the solitonic, breather and moving droplet phase for the velocities $v=0, 0.05$, and $0.5$, respectively (see Fig.~\ref{fig:finiteg2}(d)-(f)). Further, to probe the effect of mean-field interaction on the transition from breathing to moving droplet in Fig.~\ref{fig:dendelgsw}, we show the dynamical evolution of the stripe droplet for various $\delta g/g$ by keeping the velocity fixed to $v=0.2$. The other parameters have been kept the same as in Fig.~\ref{fig:finiteg2}. At $\delta g/g=0$, the stripe droplet exhibits the breathing-like oscillation with period $T\sim 500$ (See Fig.~\ref{fig:dendelgsw}(a)) which increases to $T\sim 1200 $ for $\delta g/g=0.1$. Upon further increase in the $\delta g/g=0.5$ results in the breaking of the droplet along with the presence of the moving droplet. The effect of $\delta g/g$ on the dynamical evolution of the droplet is similar to those of a bright droplet. We compute the critical velocity at which the transition from the breathing droplet to the moving droplet occurs for the stripe droplet region by analyzing the velocity at which the chemical potential attains the minimum~\cite{Gangwar2022}. In Fig.~\ref{fig:chemPot1}, we show the variation of the $\mu$ with $v$ for different $\delta g/g$. The $v_c$ are $\sim 0.40$, $\sim 0.28$ and $\sim 0.14$ respectively for $\delta g/g=0, 0.2$ and $0.5$. To make the analysis more complete in Fig.~\ref{fig:phaseplotsw}, we illustrate the breathing and moving stripe droplet regime with the blue dots and red diamonds, respectively, in the phase space $\delta g/g$ and $v$. We calculate the critical velocity by analyzing the minima of corresponding chemical potential and show those boundary regions with the dotted line in the Fig. We notice that the critical velocity for the transition decreases with the increase in $\delta g/g$ similar to those observed for the bright droplet regime. However, we find that the critical velocity for a given $\delta g/g$ is lower for the stripe droplet regime than that for the bright droplets.


Having analyzed the dynamics of the quantum droplet in detail by perturbing the condensate through a velocity, in the following, we investigate different dynamics by suddenly quenching the interaction strength.

\subsection{Quench dynamics of quantum droplets}
\label{sec:3c}
Recently Gangwar et al. carried out an extensive numerical simulation to analyze the effect of the quenching of coupling parameters on the overall dynamics of the bright and stripe phase of the quantum droplet with $\delta g=0$. They reported a large variety of novel features, like the dynamical phase transition from plane to the stripe soliton, dynamical flipping of the spin state, generation of the secondary solitons, etc., by quenching, either the Rabi coupling or SO coupling~\cite{Gangwar2022}. Following this, we analyze the effect of the interaction strength quenching on the dynamics of the droplet for both zero and finite $\delta g/g$.

\begin{figure*}[!htp]
\centering\includegraphics[width=0.99\linewidth]{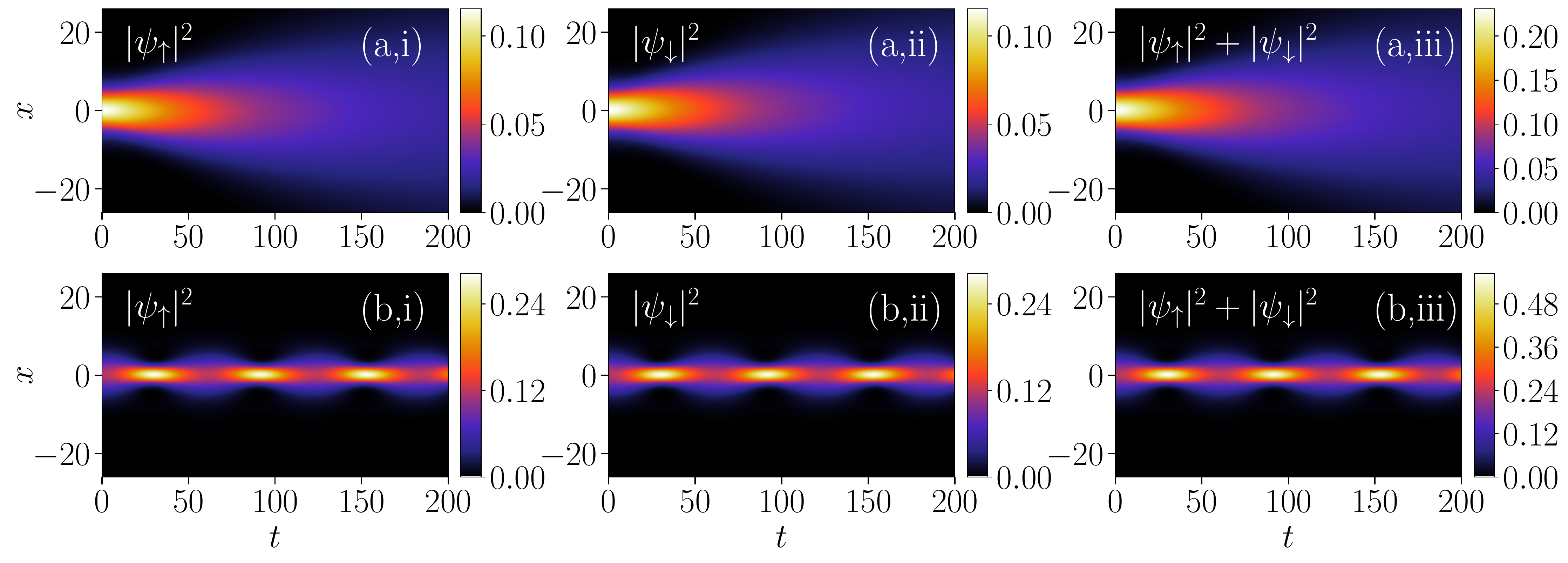} 
\caption{Dynamics of quantum droplet appearing due to sudden quenching in interaction strength as the ground states are prepared for $\Omega =1$, $k_L = 0.5$, $g = -g_{\uparrow\downarrow}=0.5$. (a,i)-(a,iii): for quenching from $g= -g_{\uparrow\downarrow}=0.5 \to g= -g_{\uparrow\downarrow}=0.1$. The quenching triggers the expansion in the droplet for a long time. (b,i)-(b,iii): for quenching from $g= -g_{\uparrow\downarrow}=0.5 \to g= -g_{\uparrow\downarrow}=1$. The quenching induces the appearance of breathing droplets.}
\label{fig:queng}
\end{figure*}


\begin{figure*}[!htp]
\centering\includegraphics[width=0.99\linewidth]{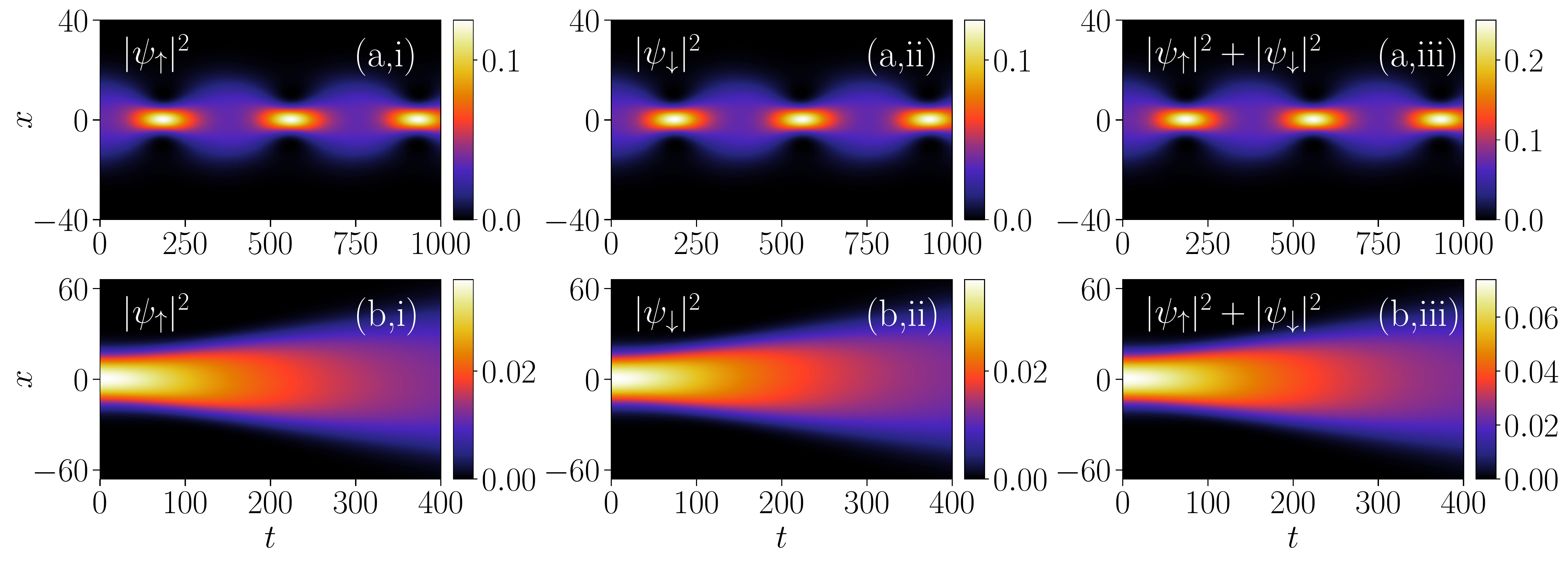} 
\caption{Quenching dynamics of the quantum droplet for the different quench protocols with the initial ground state prepared at $\Omega =1$, $k_L = 0.5$, $g=0.5$, $g_{\uparrow\downarrow}=-0.25$ $(\delta g/g = 0.5)$. (a,i)-(a,iii): When the inter-species interaction strength is quenched as $g_{\uparrow\downarrow}=-0.25\to -0.45$, $(\delta g/g = 0.5 \to 0.1)$ breathing droplets are observed. (b,i)-(b,iii): quenching $g_{\uparrow\downarrow}=-0.25\to -0.05$ $(\delta g/g = 0.5 \to 0.9)$ results the droplets expansion after $t\sim 100$.}
\label{fig:quendelgbyg}
\end{figure*}


Firstly, we analyze the quench dynamics by applying sudden changes in the interaction strength in both cases, $\delta g/g=0$ and $\delta g/g \ne 0$. For this purpose, we prepare the ground state of the quantum droplet by fixing $g=-g_{\uparrow \downarrow}=0.5$ and $\Omega=1$, $k_L=0.5$. As we make a sudden change in the interaction strength from higher to lower interaction ($g =0.5 \to 0.1$), the QD shows an expansion immediately after the quenching [see Fig.~\ref{fig:queng}(a,i)-(a,iii)]. However, when we quench from lower to higher interaction ($g=0.5 \to 1$), the quantum droplet starts exhibiting breather-like oscillations as shown in Fig.~\ref{fig:queng}(b,i)-(b,iii). This particular feature of the quenched behaviour of the QD demonstrate controlling the nature of the droplet from the expanding to the breathers by simply changing the quenching protocols.

In Fig.~\ref{fig:quendelgbyg}, we show the dynamical behaviour of the quantum droplet ensuing due to the quenching of interspecies interaction strength ($g_{\uparrow \downarrow}$). Like the case with $\delta g/g=0$, here we also consider two types of quenching protocols, from large $g_{\uparrow \downarrow}$ to small and vice versa. We generate a ground state by fixing the interaction strength $g=0.5$, $g_{\uparrow \downarrow}=-0.25$ ($\delta g/g=0.5$), $\Omega=1$ and $k_L=0.5$. Upon quenching from a small $g_{\uparrow \downarrow}$ ($g_{\uparrow \downarrow}=-0.25$) to a large $g_{\uparrow \downarrow}$ ($g_{\uparrow \downarrow}=-0.45$), triggers the breathers like oscillation as shown in Fig.~\ref{fig:quendelgbyg}(a,i)-(a,iii). This particular feature can be attributed to the fact that upon increasing the attractive interspecies interaction may lead to domination of LHY on the mean-field which may result to these kinds of breathing behaviour~\cite{Mistakidis2021}. In the reverse case, when we quench $g_{\uparrow \downarrow} = -0.25 \to -0.05$, the QDs undergo an expansion after $t=100$ [see Fig.~\ref{fig:quendelgbyg}(b,i)-(b,iii)] which can be related to the domination of the mean-field term over LHY term~\cite{Mistakidis2021}. As we compare the quench dynamics that we observe for $\delta g/g=0$ to those for finite $\delta g/g$, we find that both breathers and expansion dynamical features depend quite strongly on the $\delta g/g$. For $\delta g/g \ne 0$, this dynamical behaviour starts at a later time than that for $\delta g/g=0$. Therefore the presence of the $\delta g/g$ can be associated with enhancing the dynamical stability of the droplet~\cite{Mithun2020}.

\begin{figure*}[!ht]
\centering\includegraphics[width=0.99\linewidth]{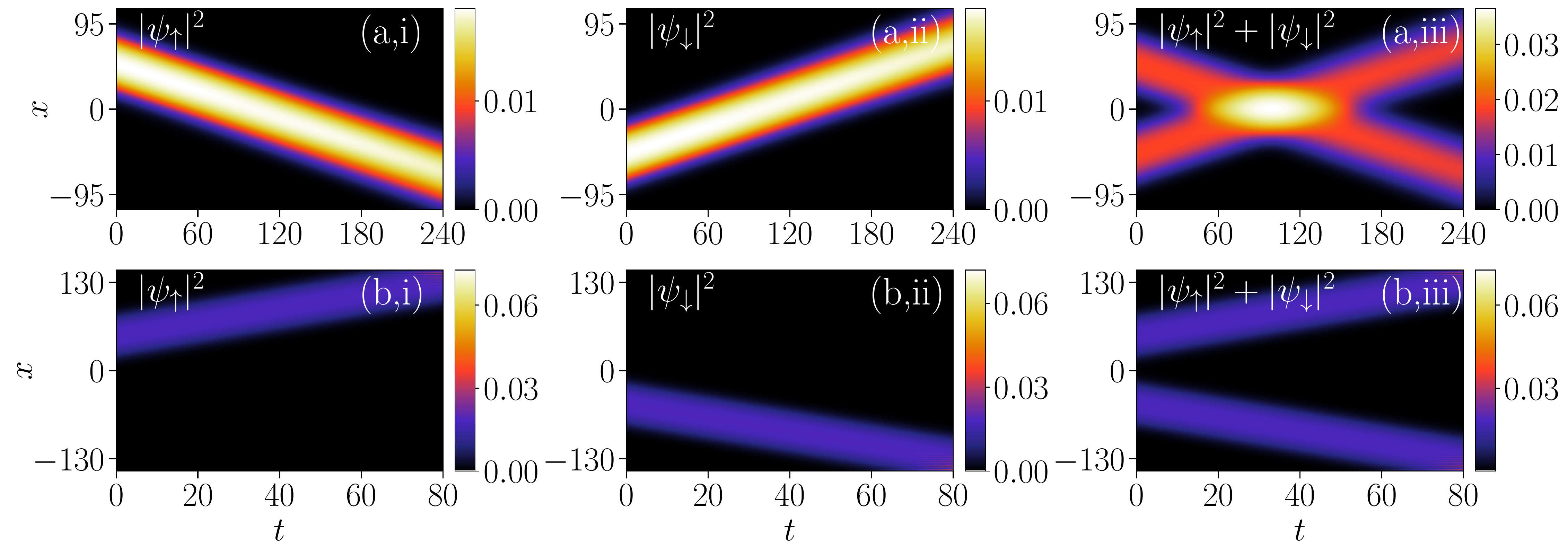}
\caption{Collisional dynamics of the quantum droplet arising due to different quenching protocols. (a,i)-(a,iii): For quenching $k_L=0 \to 0.5$, the droplet undergoes elastic collision and (b,i)-(b,iii): for quenching $k_L=0 \to 2$, the droplets appear to repel each other and move in the opposite direction. The ground state was prepared with $g=0.5, g_{\uparrow\downarrow}=-0.25$, $\Omega=0$ and $k_L=0$ and subsequently equal and opposite velocities with magnitude $v=1$ was given to the individual components at $t=0$.}
\label{fig:collision-quench1}
\end{figure*}

Now we focus on analyzing the collision dynamics of the droplet for both $\delta g/g=0$ and $\delta g/g \ne 0$. Previously Gangwar et al. presented an extensive collision dynamics of the droplet for $\delta g=0$ and found the presence of a variety of elastic and inelastic collisions of the droplet depending upon their velocities. Following this, here, we discuss the effect of the $\delta g/g$ on the collision dynamics that arises either by quenching the SO coupling parameter or by providing equal and opposite velocities to both components. 

\begin{figure*}[!htp]
\centering\includegraphics[width=0.99\linewidth]{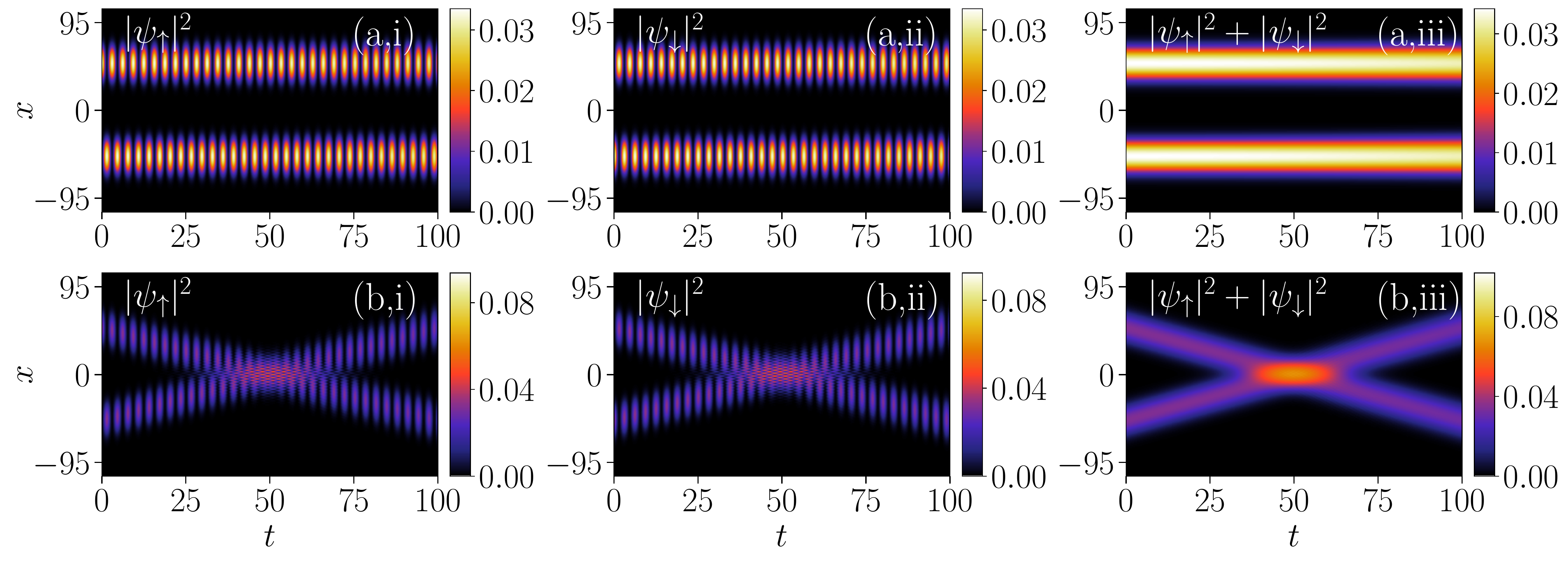} 
\caption{Collisional dynamics of the quantum droplet prepared with $g=-g_{\uparrow \downarrow}=0.5$ and $\Omega=1, k_L=0$. (a,i-a,iii) $v=0$ and (b,i-b,iii) $v=1$. For $v=0$, the droplet displays spin-flipping-like behaviour with no motion and shows expansion after $t \sim 50$. However, for $v=1$, the droplet exhibits a space-time breather with occurrence of an inelastic collision around $t \sim 50$.}
\label{fig:colli1}
\end{figure*}


In Fig.~\ref{fig:collision-quench1}, we show the collision dynamics for the droplet with the ground state prepared with $g=0.5$, $g_{\uparrow \downarrow}=-0.25$ ($\delta g/g=0.5$), and $\Omega=k_L=0$. Further, the components were kept at $\pm 50$ and were given the initial velocity $v=1$ in the $-x$ (up-component) and $+x$ (down-component) direction. Following this, at $t=0$ SO coupling was quenched from $k_L=0 \to 0.5$ that resulting in the elastic collision between the droplets around $t \sim 80$ as shown in Fig.~\ref{fig:collision-quench1}(a,i)-(a,iii). However, upon quenching the $k_L =0 \to 2$ results in a change in the direction of the motion of the components leading to their separation at a longer time [see Fig.~\ref{fig:collision-quench1}(b,i)-(b,iii)]. The different features of the collision indicate that the collisional dynamics could be controlled by quenching strength of the SO coupling parameters. Interestingly for finite $\delta g/g$ the collision is elastic in nature while it is inelastic in nature for $\delta g =0$ for the same SO quenching protocol~\cite{Gangwar2022}.   

\begin{figure*}[!htp]
\centering\includegraphics[width=0.49\linewidth]{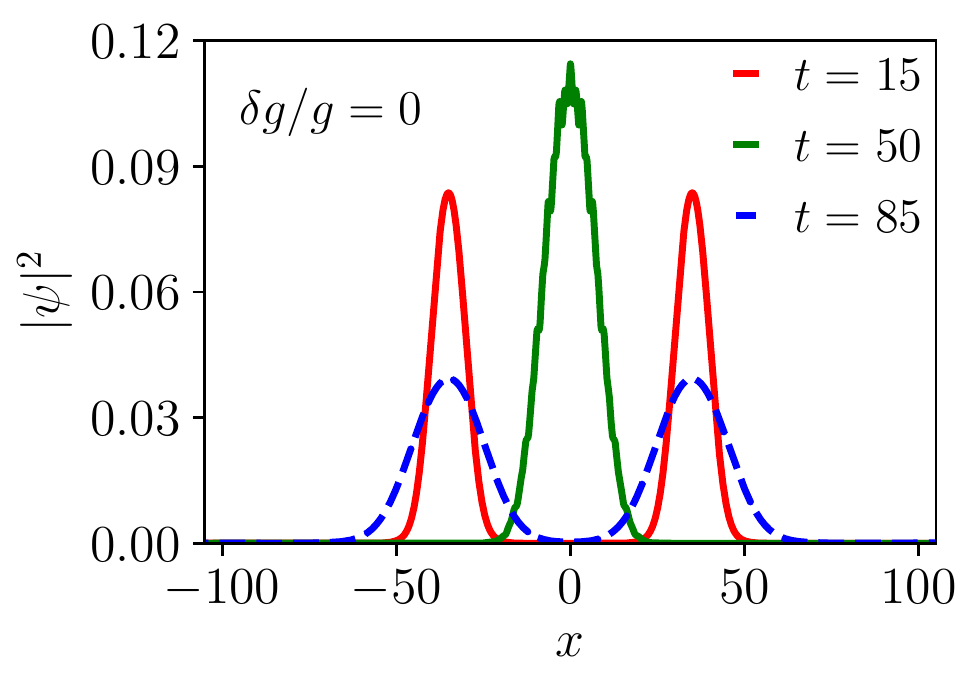} 
\caption{Total density plot at different instant of time for the QD collision as shown in the Fig.~\ref{fig:colli1}: at $t=15$ (before collision), at $t=50$ (during the collision) and at $t=85$ (after the collision). The change in the amplitude and expansion of the density after the collision indicate the inelastic nature of the collision. }
\label{fig:colli1Den}
\end{figure*}


\begin{figure*}[!htp]
\centering\includegraphics[width=0.99\linewidth]{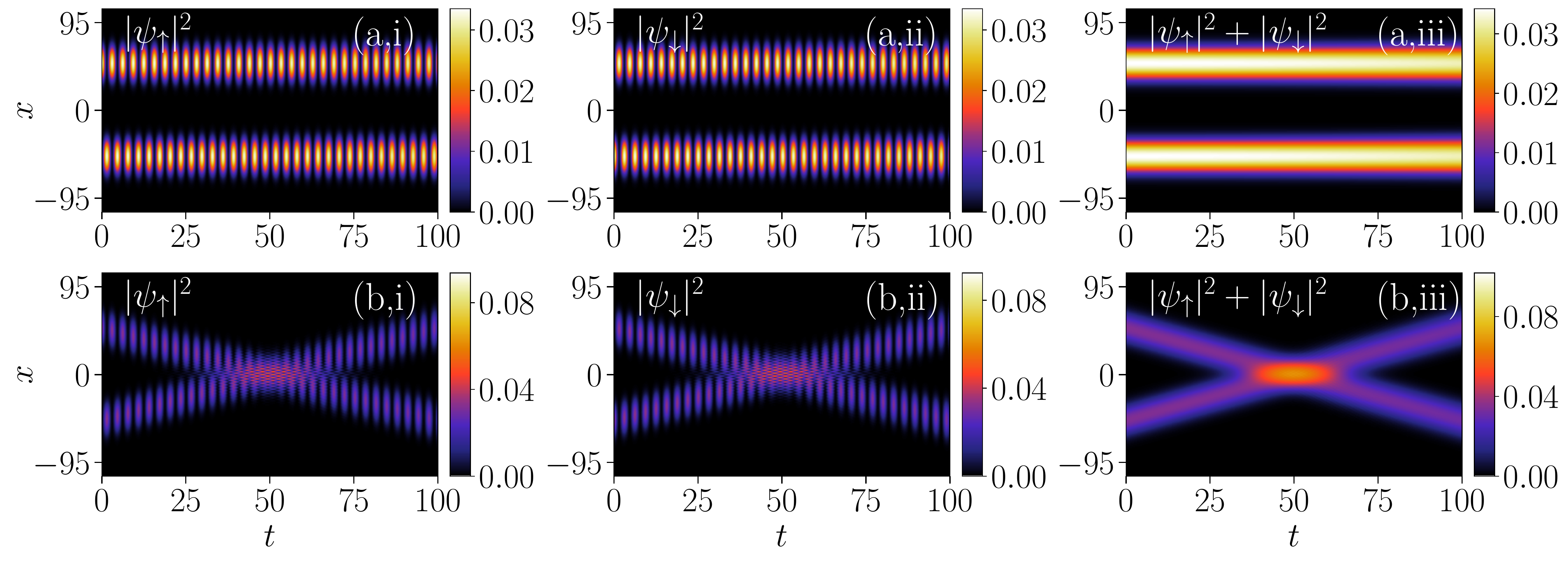} 
\caption{Collisional dynamics of the quantum droplet prepared with $g=0.5$, $g_{\uparrow \downarrow}=-0.25$ ($\delta g/g=0.5$) and $\Omega=1$, $k_L=0$. (a,i)-(a,iii) For $v=0$ droplet exhibits spin-flipping without expansion and (b,i)-(b,iii) for $v=1$ droplet undergoes elastic collision at $t \sim 50$. }
\label{fig:colli2}
\end{figure*}


Now we discuss the collisional dynamics for both $\delta g/g=0$ and $\delta g/g \ne 0$ with finite Rabi coupling $\Omega=1$ and $k_L=0$. In Fig.~\ref{fig:colli1}, we show the collisional dynamics for $\delta g/g=0$ with $g=-g{\uparrow \downarrow} =0.5$ for zero and finite ($v=1$) velocity of the droplet. In Fig.~\ref{fig:colli1}(a,i)-(a,iii) we illustrate the evolution of the droplet for $v=0$. For this case the droplets are positioned at $x=\pm 50$. With time the droplet does not move from its initial position, however, the individual components show some sort of breathing like behaviour associated with spin-flipping-like dynamics. After some time ($t \sim 25$) the droplet undergoes expansion. Similar type of spin-flipping has also been realized for SO coupled BECs by making an atom number imbalance at the initial time~\cite{Cheng2014}.

As we consider the dynamics of the droplet by giving the velocity $v=1$ to the individual components we find that the droplet undergoes inelastic collision at $t \sim 50$ [see Fig.~\ref{fig:colli1}(b,i)-(b,iii)]. To understand the nature of the collision more clearly in Fig.~\ref{fig:colli1Den} we show the total density of the droplet before ($t=15$), during ($t=50$) and after ($t=85$) the collision. The total density of the droplet shows expansion after the collision complementing the inelastic nature of the collision~\cite{Young2022}. 

\begin{figure*}[!htp]
\centering\includegraphics[width=0.49\linewidth]{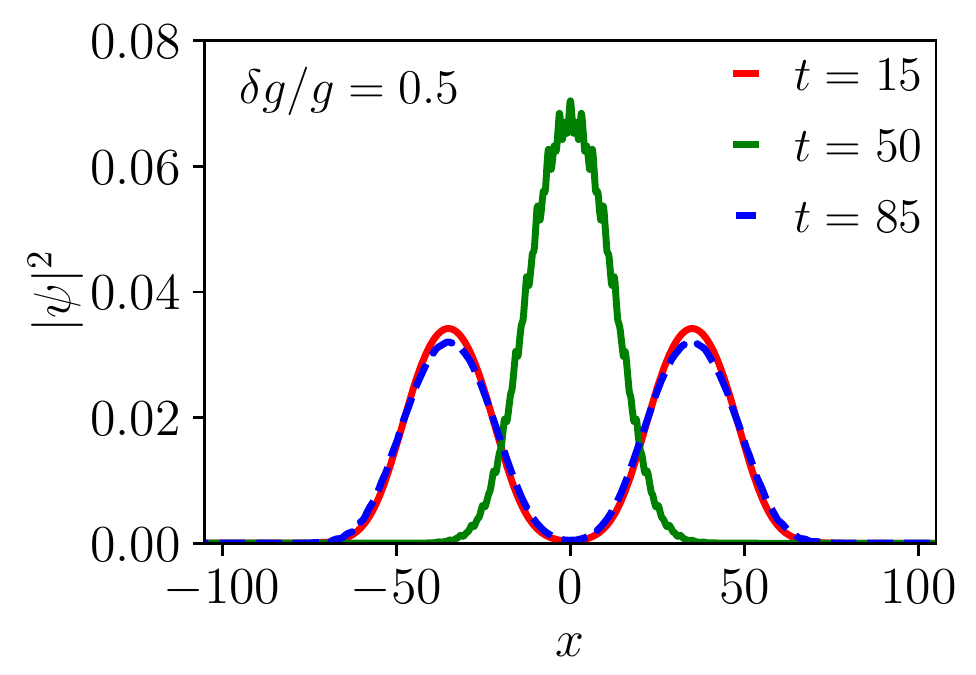} 
\caption{Total density plot at different instant of time for the QD collision as shown in the Fig.~\ref{fig:colli2}: before ($t=15$), during ($t=50$) and after the collision ($t=85$). The density remains unchanged after the collision inferring the elastic nature of the collision in presence of finite $\delta g/g$.}
\label{fig:colli2Den}
\end{figure*}


Further to probe the role of the finite contribution of the mean-field term on the collisional dynamics as discussed above we change the $\delta g/g=0$ to $\delta g/g=0.5$ by keeping the other parameters same as those we have in Fig.~\ref{fig:colli1} and plot the time evolution of the individual components as well the total density in the Fig.~\ref{fig:colli2}(a,i-a,iii) for $v=0$ and Fig.~\ref{fig:colli2}(b,i-b,iii) for $v=1$. We find that the breathing along with the spin-flipping dynamics of the component remains same as those we observed for $\delta g/g=0$ for both the velocities. However, droplets appear to be more stable for $\delta g/g=0.5$. For instance at $v=0$ the droplets exhibit spin-flipping dynamics without any expansion for longer time [See Fig.~\ref{fig:colli2}(a,i-a,iii)]. However, for $v=1$ the droplet appears to undergo elastic collision at $t\sim 50$. The elastic nature of the collision has been further established by analyzing the nature of the density of the droplet before and after the collision as shown in the Fig.~\ref{fig:colli2Den}. Overall we find that the presence of finite mean-field term enhances the overall dynamical stability of the droplet~\cite{Mithun2020}.

\section{Summary and Conclusions}
\label{sec:4}
In this paper, we have systematically analyzed the effect of the finite mean-field term on the structure and dynamics of different phases of the quantum droplet, namely, quantum bright droplet and quantum stripe droplet. We find that the density of both bright droplet and stripe droplet phases decrease with enlargement in its size upon increase in the mean-field contribution. Further we have analyzed the effect of velocity perturbation on the dynamics of the droplet. Increasing the strength of the velocity perturbation to the initial ground state results the appearance of breather droplets for low velocity, which transforms into the moving droplet at high velocities. The critical velocity at which the transition from the breather to the moving droplets occurs decreases upon increasing the SO coupling parameters for both $\delta g/g=0$ and $\delta g/g \ne 0$. Next we have analyzed the quenching dynamics resulting from the quenching of the interspecies nonlinear interactions. We find that as we perform quenching from high to low nonlinear interaction, droplets exhibit breather-like oscillations. However, these breather droplet gets transformed into expanding one as the quenching is executed in the interspecies interaction from low to high values. The behavior of the droplet in the presence of $\delta g/g$ was observed to be dynamically more stable than those for $\delta g/g=0$. Finally, we have analyzed the collisional dynamics of the droplets by giving equal and opposite velocities to the component and found that while the collision is inelastic for $\delta g/g =0 $, it is elastic for $\delta g/g \ne 0$.      






\section{Acknowledgments}
R.R. acknowledges support from the Ministry of Science and Technology (MOST), Taiwan, under Grant No. MOST-111-2119-M-001-002. P.K.M. acknowledges the Department of Science and Technology - Science and Engineering Research Board (DST-SERB) India for the financial support through Project No. ECR/2017/002639. The work of P.M. is supported by DST-SERB under Grant No. CRG/2019/004059, DST-FIST under Grant No. SR/FST/PSI-204/2015(C), and MoE RUSA 2.0 (Physical Sciences).

\end{document}